# Normalization of zero-inflated data:
# An empirical analysis of a new indicator family

Robin Haunschild[1]   Lutz Bornmann[2]

[1] *r.haunschild@fkf.mpg.de*
Max Planck Institute for Solid State Research, Heisenbergstr. 1, 70569 Stuttgart (Germany)

[2] *bornmann@gv.mpg.de*
Division for Science and Innovation Studies, Administrative Headquarters of the Max Planck Society, Hofgartenstr. 8, 80539 Munich (Germany)

**Abstract**
Recently, two new indicators (Equalized Mean-based Normalized Proportion Cited, EMNPC, and Mean-based Normalized Proportion Cited, MNPC) were proposed which are intended for sparse data. We propose a third indicator (Mantel-Haenszel quotient, MHq) belonging to the same indicator family. The MHq is based on the MH analysis – an established method for polling the data from multiple 2×2 cross tables based on different subgroups. We test (using citations and assessments by peers, i.e. F1000Prime recommendations) if the three indicators can distinguish between different quality levels as defined on the basis of the assessments by peers (convergent validity). We find that the indicator MHq is able to distinguish between the quality levels in most cases while MNPC and EMNPC are not.

**Conference Topic**
Indicators

**Introduction**
Alternative metrics (altmetrics) have been established as a new fast-moving and dynamic area in scientometrics (Galloway, Pease, & Rauh, 2013). Initially, altmetrics have been proposed as an alternative to traditional bibliometric indicators. Altmetrics are a collection of multiple digital indicators which measure activity related to research papers on social media platforms, in mainstream media or in policy documents (National Information Standards Organization, 2016; Work, Haustein, Bowman, & Larivière, 2015). According to Haustein (2016), sources of altmetrics can be grouped into (i) social networks, (ii) social bookmarks and online reference management, (iii) social data (e.g., data sets, software, presentations), (iv) blogs, (v) microblogs, (vi) wikis, and (vii) recommendations, ratings, and reviews.

Recently, some indicators based on altmetrics have been proposed which are normalized with respect to the scientific field and publication year. These indicators were developed because studies have shown that altmetrics are – similar to bibliometric data – field- and time-dependent (see, e.g., Bornmann, 2014). Some fields are more relevant to the general public or a broader audience than other fields (Haustein, Larivière, Thelwall, Amyot, & Peters, 2014). The Mean Normalized Reader Score (MNRS) was introduced by Haunschild and Bornmann (2016) for normalization of data from social bookmarks and online reference management platforms (with a special emphasis on Mendeley readers), see also Fairclough and Thelwall (2015). The Mean Discipline Normalized Reader Score (MDNRS) was tailored specifically to Mendeley by Bornmann and Haunschild (2016b). The MNDRS uses Mendeley disciplines for field normalization. The employed normalization procedures rely on average value calculations across scientific fields and publication years as expectation values. Normalization procedures based on averages and percentiles of individual papers are problematic for zero-inflated data sets (Haunschild, Schier, & Bornmann, 2016). Bornmann and Haunschild (2016a) proposed the Twitter Percentile (TP) – a field- and time-normalized indicator for Twitter data.

The overview of Work, et al. (2015) on studies investigating the coverage of papers on social media platforms show that many platforms have coverages of less than 5% (e.g., Blogs, or Wikipedia). Erdt, Nagarajan, Sin, and Theng (2016) report similar findings in their meta-analysis. They found that former empirical studies dealing with the coverage of altmetrics show that about half of the platforms are at or below 5%; except for three (out of eleven) where the coverage is below 10%. Thus, altmetrics data are frequently concerned by zero-inflation.

Bornmann and Haunschild (2016a) circumvented the problem of zero-inflated Twitter data by including in the TP calculation only journals with at least 80% of the papers having at least 1 tweet each. However, this procedure leads to the exclusion of many journals from the TP procedure. Very recently, Thelwall (2017a, 2017b) proposed a new family of field- and time-normalized indicators for zero-inflated altmetrics data. The new indicators family is based on units of analysis (e.g., a researcher or institution) rather than on the paper level. They compare the proportion of mentioned papers (e.g., on Wikipedia) of a unit with the proportion of mentioned papers in the corresponding fields and publication years (the expected values). The family consists of the Equalized Mean-based Normalized Proportion Cited (EMNPC) and the Mean-based Normalized Proportion Cited (MNPC). Hitherto, this new family of indicators has only been studied on rather small samples.

In this study, we investigate the new indicator family empirically on a large scale (multiple complete publication years) and add a further indicator to this family. In statistics, the Mantel-Haenszel (MH) analysis is recommended for polling the data from multiple 2×2 cross tables based on different subgroups (here: mentioned and not-mentioned papers of a unit published in different subject categories and publication years compared with the corresponding reference sets). We call the new indicator Mantel-Haenszel quotient (MHq). In the empirical analysis, we compare the indicator scores with ratings by peers. We investigate whether the indicators are able to discriminate between different quality levels assigned by peers to publications. Thus, we test the convergent validity of the indicators. Since the convergent validity can only be tested by using citations (which are related to quality), the empirical part is based solely on citations and not altmetrics data. Good performance on the convergent validity test is a necessary condition for the use of the indicators in altmetrics (although for altmetrics, the relationship to quality is not clear).

**Indicators for zero-inflated count data**

Whereas the EMNPC and MNPC proposed by Thelwall (2017a, 2017b) are explained in the following two sections, the MHq is firstly introduced in the section thereafter. The next sections present not only the formulas for the calculation of the three metrics, but also the corresponding 95% confidence intervals (CIs). The CI is a range of possible indicator values: We can be 95% confident that the interval includes the "true" indicator value in the population. With the use of CIs, we assume that we analyse sample data and infer to a larger, inaccessible population (Williams & Bornmann, 2016). According to Claveau (2016) the general argument for using inferential statistics with scientometric data is "that these observations are realizations of an underlying data generating process … The goal is to learn properties of the data generating process. The set of observations to which we have access, although they are all the actual realizations of the process, do not constitute the set of all possible realizations. In consequence, we face the standard situation of having to infer from an accessible set of observations – what is normally called the sample – to a larger, inaccessible one – the population. Inferential statistics are thus pertinent" (p. 1233).

The relationship between CIs and statistical significance (in case of independent proportions) is as follows:

"1. If the 95% CIs on two independent proportions just touch end-to-end, overlap is zero and the p value for testing the null hypothesis of no difference is approximately .01.
2. If there's a gap between the CIs, meaning no overlap, then p<.01.
3. Moderate overlap … of the two CIs implies that p is approximately .05. Less overlap means p<.05.
Moderate overlap is overlap of about half the average length of the overlapping arms" (Cumming, 2012, p. 402).

*Equalized Mean-based Normalized Proportion Cited (EMNPC)*

Thelwall (2017a, 2017b) introduced the EMNPC as an alternative indicator for zero-inflated count data. The approach of the EMNPC is to calculate the proportion of papers that are mentioned: suppose that publication set $g$ has $n_{gf}$ papers in the publication year and subject category combination $f$. $s_{gf}$ of the papers are mentioned (e.g., on Wikipedia). $F$ is defined as all publication year and subject category combinations of the papers in the set. The overall proportion of $g$'s papers that are mentioned is the number of mentioned papers ($s_{gf}$) divided by the total number of papers ($n_{gf}$):

$$p_g = \sum_{f \in F} s_{gf} \Big/ \sum_{f \in F} n_{gf} \qquad (1)$$

However, $p_g$ could lead to misleading results if the publication set $g$ includes many papers which are published in fields with many mentioned papers. Thelwall (2017a, 2017b) proposes to avoid the problem by artificially treating $g$ as having the same number of papers in each publication year and subject category combination. The author fixes it to the arithmetic average of numbers in each combination, but recommends not including in the analysis combinations of $g$ with only a few papers. Thus, the equalized sample proportion of $g$, $\hat{p}$ is the simple average of the proportions in each combination

$$\hat{p}_g = \frac{\sum_{f \in F} \frac{s_{gf}}{n_{gf}}}{[F]} \qquad (2)$$

The corresponding world sample proportion is defined as:

$$\hat{p}_w = \frac{\sum_{f \in F} \frac{s_{wf}}{n_{wf}}}{[F]} \qquad (3)$$

In Eqns. (2) and (3), [F] is the number of subject category and publication year combinations in which the group (in case of Eq. (2)) and the world (in case of Eq. (3)) publishes. Thus, the equalized group sample proportion has the undesirable property that it treats $g$ as if the average mentions of its papers do not vary between the subject categories. The EMNPC for each publication set $g$ is the ratio of both equalized sample proportions:

$$\text{EMNPC} = \hat{p}_g / \hat{p}_w \qquad (4)$$

CIs for the EMNPC can be calculated as follows (Thelwall, 2017a):

$$EMNPC_L = \exp\left(\ln\left(\frac{\hat{p}_g}{\hat{p}_w}\right) - 1.96 \sqrt{\frac{(n_g - \hat{p}_g n_g)/(\hat{p}_g n_g)}{n_g} + \frac{(n_w - \hat{p}_w n_w)/(\hat{p}_w n_w)}{n_w}}\right) \qquad (5)$$

$$EMNPC_U = \exp\left(\ln\left(\frac{\hat{p}_g}{\hat{p}_w}\right) + 1.96\sqrt{\frac{(n_g-\hat{p}_g n_g)/(\hat{p}_g n_g)}{n_g} + \frac{(n_w-\hat{p}_w n_w)/(\hat{p}_w n_w)}{n_w}}\right) \quad (6)$$

Here, $n_g$ is the total sample size of the group and $n_w$ is the total sample size of the world.

*Mean-based Normalized Proportion Cited (MNPC)*

The second indicator proposed by Thelwall (2017a) has been named as Mean-based Normalized Proportion Cited (MNPC). The MNPC is calculated as follows: For each paper with at least one mention (e.g., on Wikipedia), the number of mentions is replaced by the reciprocal of the world proportion mentioned for the corresponding subject category and publication year. All other papers with zero mentions remain at zero. Let $p_{gf} = s_{gf}/n_{gf}$ be the proportion of papers mentioned for publication set $g$ in the corresponding subject category and publication year combination $f$ and let $p_{wf} = s_{wf}/n_{wf}$ be the proportion of world's papers cited in the same year and subject category combination $f$. Then:

$$r_i = \begin{cases} 0, & \text{if } c_i = 0 \\ 1/p_{wf} & \text{if } c_i > 0, \text{ where paper } i \text{ is from year and subject category combination } f \end{cases} \quad (7)$$

Following the calculation of the MNCS (Waltman, van Eck, van Leeuwen, Visser, & van Raan, 2011), the MNPC is defined as

$$MNPC = \frac{(r_1 + r_2 + \cdots r_n)}{n} \quad (8)$$

An approximate CI has been constructed by Thelwall (2016, 2017a) for the MNPC. In the first step, the lower limit $L$ (MNPC$_{fgL}$) and upper limit $U$ (MNPC$_{fgU}$) for group $g$ in subject category and publication year combination $f$ is calculated with:

$$MNPC_{gfL} = \exp\left(\ln\left(\frac{\hat{p}_{gf}}{\hat{p}_{wf}}\right) - 1.96\sqrt{\frac{(n_{gf}-\hat{p}_{gf}n_{gf})/(\hat{p}_{gf}n_{gf})}{n_{gf}} + \frac{(n_{wf}-\hat{p}_{wf}n_{wf})/(\hat{p}_{wf}n_{wf})}{n_{wf}}}\right) \quad (9)$$

$$MNPC_{gfU} = \exp\left(\ln\left(\frac{\hat{p}_{gf}}{\hat{p}_{wf}}\right) + 1.96\sqrt{\frac{(n_{gf}-\hat{p}_{gf}n_{gf})/(\hat{p}_{gf}n_{gf})}{n_{gf}} + \frac{(n_{wf}-\hat{p}_{wf}n_{wf})/(\hat{p}_{wf}n_{wf})}{n_{wf}}}\right) \quad (10)$$

In the second step, the group-specific lower and upper limits are used to calculate the MNPC CIs:

$$MNPC_L = MNPC - \sum_{f \in F} \frac{n_{gf}}{n_g}\left(\frac{p_{gf}}{p_{wf}} - MNPC_{gfL}\right) \quad (11)$$

$$MNPC_U = MNPC + \sum_{f \in F} \frac{n_{gf}}{n_g}\left(MNPC_{gfU} - \frac{p_{gf}}{p_{wf}}\right) \quad (12)$$

The MNPC cannot be calculated, if any of the world proportions are equal to zero. Furthermore, CIs cannot be calculated if any of the group proportions are equal to zero. Thus, Thelwall (2017a) proposed to remove the corresponding subject category publication year combination from the data or to add a continuity correction of 0.5 to the number of mentioned and not mentioned papers in these cases. We prefer the latter (to add 0.5 to the number of papers mentioned and not mentioned, respectively). This approach is recommended by Plackett (1974) for the calculation of odds ratios.

*Mantel-Haenszel quotient (MHq)*

For polling the data from multiple 2×2 cross tables based on different subgroups (which are part of a larger population), the most commonly used and recommended method is the MH analysis (Hollander & Wolfe, 1999; Mantel & Haenszel, 1959; Sheskin, 2007). According to Fleiss, Levin, and Paik (2003) the method "permits one to estimate the assumed common odds ratio and to test whether the overall degree of association is significant. Curiously, it is not the odds ratio itself but another measure of association that directly underlies the test for overall association … The fact that the methods use simple, closed-form formulas has much to recommend it" (p. 250). Radhakrishna (1965) demonstrate that the MH approach is formally and empirically valid against the background of clinical trials.

The MH analysis results in a summary odds ratio for multiple 2×2 cross tables which we call MHq. For the impact comparison of units in science with reference sets (the world), the 2×2 cross tables (which are polled) consist of the number of papers mentioned and not mentioned in subject category and publication year combinations *f*. Thus, in the 2×2 subject- and year-specific cross table with the cells $a_f$, $b_f$, $c_f$, and $d_f$ (see Table 1), $a_f$ is the number of mentioned papers in subject category and publication year *f*, $b_f$ is the number of not mentioned papers in subject category and publication year *f*, $c_f$ is the number of mentioned papers published by group *g* in subject category and publication year *f*, $d_f$ is the number of not-mentioned papers published by group *g* in subject category and publication year *f*. Note that the papers of group *g* are part of the papers in the world.

Table 1. 2×2 subject-specific cross table

|  | *Number of mentioned papers* | *Number of not mentioned papers* |
|---|---|---|
| Group *g* | $a_f$ | $b_f$ |
| World | $c_f$ | $d_f$ |

We start by defining some dummy variables for the MH analysis:

$$R_f = \frac{a_f d_f}{n_f} \text{ and } R = \sum_{f=1}^{F} R_f, \quad (13)$$

$$S_f = \frac{b_f c_f}{n_f} \text{ and } S = \sum_{f=1}^{F} S_f, \quad (14)$$

$$P_f = \frac{a_f + d_f}{n_f} \text{ and } Q_f = 1 - P_f \quad (15)$$

Where: $n_f = a_f + b_f + c_f + d_f$

The MHq is simply:

$$\text{MHq} = \frac{R}{S} \quad (16)$$

The CIs for MHq are calculated following Fleiss, et al. (2003). The variance of ln MHq is estimated by:

$$\widehat{Var}[\ln(MHq)] = \frac{1}{2} \left\{ \frac{\sum_{f=1}^{F} P_f R_f}{R^2} + \frac{\sum_{f=1}^{F}(P_f S_f + Q_f R_f)}{RS} + \frac{\sum_{f=1}^{F} Q_f S_f}{S^2} \right\} \quad (17)$$

The confidence interval for the MHq can be constructed with

$$MHq_L = \exp\left[\ln(MHq) - 1.96\sqrt{\widehat{Var}[\ln(MHq)]}\right] \quad (18)$$

$$MHq_U = \exp\left[\ln(MHq) + 1.96\sqrt{\widehat{Var}[\ln(MHq)]}\right] \tag{19}$$

Similar to the EMNPC and MNPC, it is an advantage of the MHq that the world average has a value of 1. It is a further advantage of the MHq that the result can be expressed as a percentage which is relative to the world average, e.g.: MHq = 1.30 means that the paper set under study has achieved an impact 30% above average.

**Data sets used**

We used the papers of the Web of Science (WoS) from our in-house database – derived from the Science Citation Index Expanded (SCI-E), Social Sciences Citation Index (SSCI), and Arts and Humanities Citation Index (AHCI) provided by Clarivate Analytics (formerly the IP and Science business of Thomson Reuters). All papers of the document type "article" with DOI published between 2010 and 2013 were included to study the indicators. Citations with a three-year citation window are retrieved from our in-house database (Glänzel & Schoepflin, 1995). For field classification, we used the overlapping WoS subject categories (Rons, 2012, 2014). In order to avoid statistical and numerical problems, we include only fields in the analysis where (1) at least 10 papers are assigned to and (2) the number of cited and uncited papers is non-zero. In total, these restrictions lead to a dataset including 4,490,998 papers.

We matched the publication data with peers' recommendations from F1000Prime. F1000Prime is a post-publication peer review system of papers from mainly medical and biological journals. Papers are selected by a peer-nominated global "Faculty" of leading scientists and clinicians who then rate the papers and explain their importance. Thus, only a restricted set of papers from the papers in these disciplines covered is reviewed, and most of the papers are actually not. The Faculty nowadays numbers more than 5,000 experts worldwide. Faculty members can choose and evaluate any paper that interests them. The papers are rated by the Faculty members as "Recommended," "Must read", or "Exceptional" which is equivalent to recommendation scores (RSs) of 1, 2, or 3, respectively. Papers can be recommended multiple times. Therefore, we calculated an average RS ($\overline{FFa}$):

$$\overline{FFa} = \frac{1}{i_{\max}}\sum_{i}^{i_{\max}} RS_i \tag{20}$$

The papers are categorized depending on their $\overline{FFa}$ value:
- Not recommended papers (Q0): $\overline{FFa} = 0$
- Recommended papers with a rather low average score (Q1): $0 < \overline{FFa} \leq 1.0$
- Recommended papers with a rather high average score (Q2): $\overline{FFa} > 1.0$

We only included fields where a paper with an F1000Prime recommendation is assigned to, following Waltman and Costas (2014). This reduces the total paper set included in the analysis to 2,873,476 papers.

**Empirical analysis**

The comparison of indicators with peer evaluation has been widely acknowledged as a way of investigating the convergent validity of metrics (Garfield, 1979; Kreiman & Maunsell, 2011). Convergent validity is the degree to which two measurements of constructs (here: two proxies of scientific quality), which should be theoretically related, are empirically related. Thelwall (2017b) justifies this approach as follows: "if indicators tend to give scores that agree to a large extent with human judgements then it would be reasonable to replace human judgements with them when a decision is not important enough to justify the time necessary for experts to read the articles in question. Indicators can be useful when the value of an assessment is not great enough to justify the time needed by experts to make human judgements" (p. 4). Several publications investigating the relationship between citations and Research Excellence Framework (REF) outcomes report considerable relationships in several subjects, such as biological science, psychology, and clinical sciences (Butler & McAllister, 2011; Mahdi, d'Este, & Neely, 2008; McKay, 2012; Smith & Eysenck, 2002; Wouters et al., 2015). Similar results were found for the Italian research assessment exercise: "The correlation strength between peer assessment and bibliometric indicators is statistically significant, although not perfect. Moreover, the strength of the association varies across disciplines, and it also depends on the discipline internal coverage of the used bibliometric database" (Franceschet & Costantini, 2011, p. 284). The overview of Bornmann (2011) shows further results on journal peer review, that a higher citation impact of papers is to be expected with better recommendations from peers.

In recent years, the correlation between the F1000Prime RSs and citation impact scores has already been targeted. The results of the regression model of Bornmann (2015) demonstrate that about 40% of publications with RS=1 belong to the 10% most frequently cited papers, compared with about 60% of publications with RS=2 and about 73% of publications with RS=3. Waltman and Costas (2014) found "a clear correlation between F1000 recommendations and citations" (p. 433). The previous results on F1000Prime allow the prognosis, therefore, that citation-based indicators differentiate more or less clearly between the three RSs. In other words, the validity of new indicators can be questioned if the ability to differentiate is not given.

Against this backdrop, we investigate in the current study the ability of the three indicators for zero-inflated count data to differentiate between the RS groups. We start with the newly introduced MHq indicator. Figure 1 shows the MHqs with CIs for the three groups (Q0, Q1, and Q2) across four publication years.

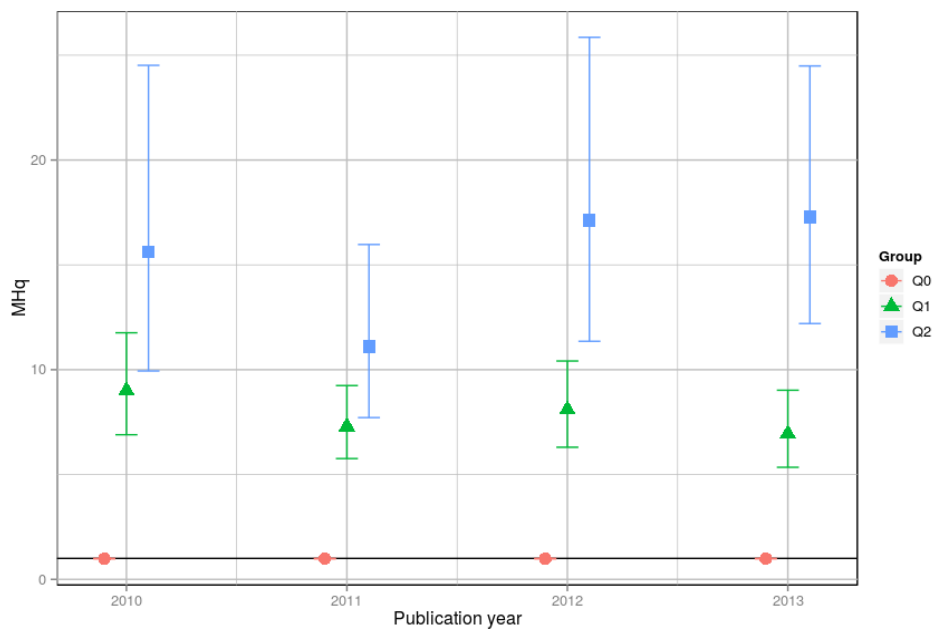

Figure 1. MHqs with CIs for the three groups and four publications years. The horizontal line with MHq=1 is the worldwide average.

It is clearly visible that the MHq values are very different for the three groups which speak for their convergent validity: The mean MHq across the years is close to (but below) 1 for Q0. The mean MHq for Q1 is about eight times and that for Q2 is about 15 times higher than the mean MHq for Q0. It seems that the MHq indicator significantly separates between the different quality levels.

However, let us take a closer look at the MHq differences between the groups on the basis of their CIs following the rules of Cumming (2012) and Cumming and Finch (2005). If there is a gap between two CIs in the figure, then the difference is statistically significant ($p<.01$). This is the case for the years 2012 and 2013. Here, the indicator differentiates clearly and statistically significantly between the three groups ($p<.01$). In 2010 and 2011, there is also a statistically significant difference between Q0 and the other two groups. However, the CIs for Q1 and Q2 overlap in 2010 and 2011. If the overlap between the CIs is less than 50%, then the difference is statistically significant on the $p<.05$ level. This rule is reasonably accurate, however, when the two margins of error (length of one arm of a CI) do not differ by more than a factor of 2. The calculation of the overlaps yields an overlap of 43% in 2010 and 57% in 2011. Thus, the difference between the MHqs is statistically not significant in 2011 ($p>.05$). Although the difference is statistically significant in 2010 ($p<.05$), we cannot assume that the rule works accurately, because the two margins of error differ by a factor of 2.1.

The reason for the better result of the MHq in 2012 and 2013 than in 2010 and 2011 might be that 2012 and 2013 contain more uncited papers than 2010 and 2011. As MHq is designed for zero-inflated count data, a better performance can be expected for 2012 and 2013.

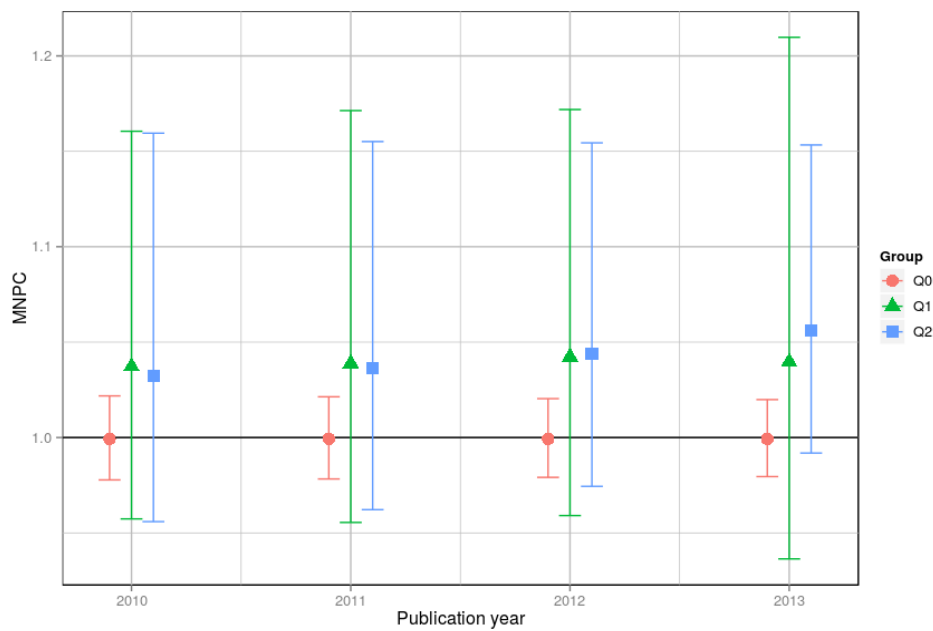

Figure 2. MNPC with CIs for the three groups and four publications years. The horizontal line with MNPC=1 is the worldwide average.

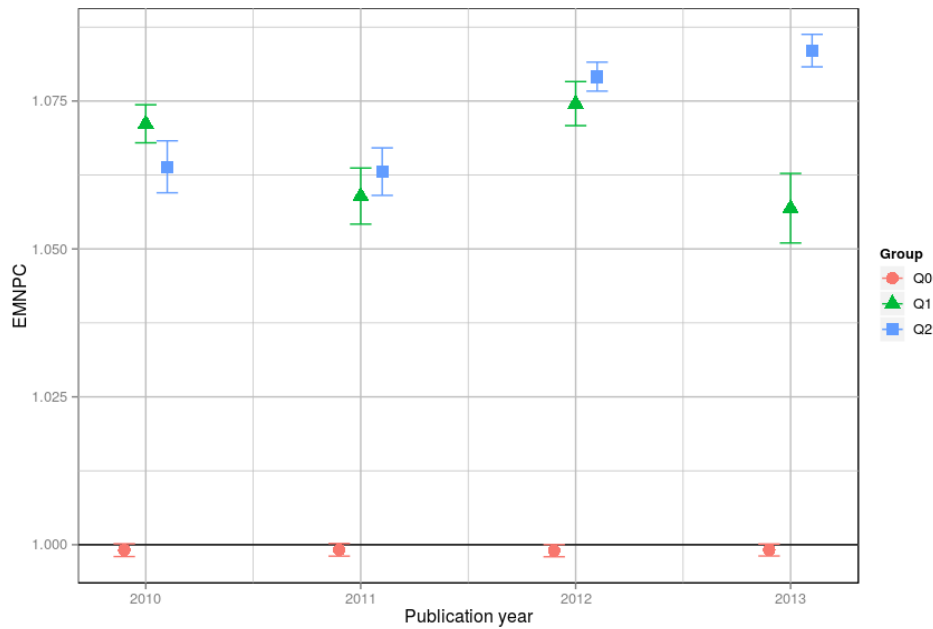

Figure 3. EMNPC with CIs for the three groups and four publications years. The horizontal line with EMNPC=1 is the worldwide average.

Figure 2 and Figure 3 show the results for the three groups for MNPC and EMNPC – the two indicators proposed by Thelwall (2017a). For both indicators, it is striking that all values in the graphs are very close to 1 – independent of the group. This is very different to Figure 1, in which the MHq values significantly differ from 1 for the two groups with recommendations (Q1 and Q2). This can be interpreted as a first sign that the MNPC and EMNPC do not differentiate between the quality levels in terms of $\overline{FFa}$ values.

The CIs in Figure 2 further reveal that the differences between the MNPCs for the different RS values are statistically not significant. There are clear and substantial overlaps for all CIs. The results in Figure 3 are very heterogeneous. In 2010, the mean value of Q2 is lower than the value of Q1. In 2013, the situation is reversed and in the expected direction then. In 2011 and 2012, the mean values are also in the expected direction, but there is a substantial overlap of the CIs (52% in 2012). According to the rules of Cumming (2012) and Cumming and Finch (2005), the differences between the CIs in in both years are statistically not significant.

**Discussion**

Although the empirical analyses in this study are based on citation data, the objective of the study is on developing indicators for sparse altmetrics data, i.e., zero-inflated altmetrics data. According to Neylon (2014), much of the data we have in altmetrics is sparse. An indicator with many zero values is unlikely to be informative about a scientific unit (e.g., a researcher or institution) in the first place (Thelwall, Kousha, Dinsmore, & Dolby, 2016). Thus, Thelwall (2017a, 2017b) proposed a new family of (meaningful) field- and time normalized indicators which are especially designed for the use with sparse data. The family consists of the EMNPC and MNPC indicators. Basically, the indicators compare the proportion of mentioned papers of a unit with the proportion of mentioned papers in the corresponding fields and publication years (the expected values).

The indicators of the family differ from most of the other indicators which have been proposed in bibliometrics and altmetrics hitherto. The other indicators are calculated for single publications and the user of the indicators can aggregate the indicator values (by averaging, summing, etc.). The indicators of the new family are not calculated for single publications, but publication sets of groups (e.g., single researchers or institutes). Thus, these indicators cannot be used as flexible as the other bibliometric and altmetric indicators. However, we think that it will never be possible to develop reliable indicators with values for single publications for zero-inflated count data.

In this study, we analyzed the new indicator family empirically and added a further indicator variant – the MHq. We did not include altmetrics data in the empirical part of the study, although the indicator family focusses on them. Before the indicators can be used with altmetrics data, they have to be validated and this can only be done on the basis of citation data. Citation data allows formulating predictions which can be empirically validated with the new indicators. In this study, we tested with citation data whether the indicators are able to differentiate validly between three quality levels – as defined by F1000 RSs ($\overline{FFa}$). Thus, we compared the indicator values with ratings by peers: Are the indicators able to discriminate between different quality levels assigned by peers to publications?

For the study, citations with a three-year citation window are retrieved from our in-house database as a compromise between having a significant correlation with quality (in the sense of post-publication peer assessments) and having a data set with rather many not mentioned papers. The results for the EMNPC and MNPC show that they cannot discriminate validly between the different quality levels. The scores for all quality levels are close to 1 (the worldwide average) and the CIs substantially overlap in many comparisons. Thus, the results point out that the convergent validity of the EMNPC and MNPC is not given. In this study, we further introduced the MHq to the new indicator family which is based on the established MH analysis. Since the MHq was able to discriminate empirically between the different quality levels – in most of the cases statistically significant – the convergent validity of the new variant seems to be given.

This study follows the important initiative of Thelwall (2017a, 2017b) to design new indicators for sparse data. Our study was the first independent attempt to investigate this indicator family empirically. Since this family is important especially for altmetrics data, we need further empirical studies which focus, e.g., on more sparse data than we used. Future empirical studies should investigate the new indicator family in other disciplines than biomedicine. F1000 focuses on the biomedical literature only.

**Acknowledgements**

The F1000Prime recommendations were taken from a data set retrieved from F1000 in November, 2017. We would like to thank Mike Thelwall for helpful correspondence regarding calculation of the CIs for MNPC and EMNPC.

**References**

Bornmann, L. (2011). Scientific peer review. *Annual Review of Information Science and Technology, 45*, 199-245.
Bornmann, L. (2014). Validity of altmetrics data for measuring societal impact: A study using data from Altmetric and F1000Prime. *Journal of Informetrics, 8*(4), 935-950. doi: http://dx.doi.org/10.1016/j.joi.2014.09.007.
Bornmann, L. (2015). Inter-rater reliability and convergent validity of F1000Prime peer review. *Journal of the Association for Information Science and Technology, 66*(12), 2415-2426.
Bornmann, L., & Haunschild, R. (2016a). How to normalize Twitter counts? A first attempt based on journals in the Twitter Index. *Scientometrics, 107*(3), 1405-1422. doi: 10.1007/s11192-016-1893-6.
Bornmann, L., & Haunschild, R. (2016b). Normalization of Mendeley reader impact on the reader- and paper-side: A comparison of the mean discipline normalized reader score (MDNRS) with the mean normalized reader score (MNRS) and bare reader counts. *Journal of Informetrics, 10*(3), 776-788.

Butler, L., & McAllister, I. (2011). Evaluating university research performance using metrics. *European Political Science, 10*(1), 44-58. doi: 10.1057/eps.2010.13.

Claveau, F. (2016). There should not be any mystery: A comment on sampling issues in bibliometrics. *Journal of Informetrics, 10*(4), 1233-1240. doi: http://dx.doi.org/10.1016/j.joi.2016.09.009.

Cumming, G. (2012). *Understanding the new statistics: effect sizes, confidence intervals, and meta-analysis*. London, UK: Routledge.

Cumming, G., & Finch, S. (2005). Inference by eye - Confidence intervals and how to read pictures of data. *American Psychologist, 60*(2), 170-180. doi: 10.1037/0003-066x.60.2.170.

Erdt, M., Nagarajan, A., Sin, S.-C. J., & Theng, Y.-L. (2016). Altmetrics: an analysis of the state-of-the-art in measuring research impact on social media. *Scientometrics*, 1-50. doi: 10.1007/s11192-016-2077-0.

Fairclough, R., & Thelwall, M. (2015). National research impact indicators from Mendeley readers. *Journal of Informetrics, 9*(4), 845-859. doi: http://dx.doi.org/10.1016/j.joi.2015.08.003.

Fleiss, J., Levin, B., & Paik, M. C. (2003). *Statistical methods for rates and proportions* (3. ed.). Hoboken, NJ, USA: Wiley.

Franceschet, M., & Costantini, A. (2011). The first Italian research assessment exercise: a bibliometric perspective. *Journal of Informetrics, 5*(2), 275-291. doi: DOI: 10.1016/j.joi.2010.12.002.

Galloway, L. M., Pease, J. L., & Rauh, A. E. (2013). Introduction to Altmetrics for Science, Technology, Engineering, and Mathematics (STEM) Librarians. *Science & Technology Libraries, 32*(4), 335-345. doi: 10.1080/0194262X.2013.829762.

Garfield, E. (1979). *Citation indexing - its theory and application in science, technology, and humanities*. New York, NY, USA: John Wiley & Sons, Ltd.

Glänzel, W., & Schoepflin, U. (1995). A Bibliometric Study on Aging and Reception Processes of Scientific Literature. *Journal of Information Science, 21*(1), 37-53. doi: Doi 10.1177/016555159502100104.

Haunschild, R., & Bornmann, L. (2016). Normalization of Mendeley reader counts for impact assessment. *Journal of Informetrics, 10*(1), 62-73. doi: 10.1016/j.joi.2015.11.003.

Haunschild, R., Schier, H., & Bornmann, L. (2016). Proposal of a minimum constraint for indicators based on means or averages. *Journal of Informetrics, 10*(2), 485-486. doi: 10.1016/j.joi.2016.03.003.

Haustein, S. (2016). Grand challenges in altmetrics: heterogeneity, data quality and dependencies. *Scientometrics*, 1-11. doi: 10.1007/s11192-016-1910-9.

Haustein, S., Larivière, V., Thelwall, M., Amyot, D., & Peters, I. (2014). Tweets vs. Mendeley readers: How do these two social media metrics differ? *it – Information Technology 2014; 56(5): , 56*(5), 207-215.

Hollander, M., & Wolfe, D. A. (1999). *Nonparametric Statistical Methods*. New York, NY, USA: Wiley.

Kreiman, G., & Maunsell, J. H. R. (2011). Nine criteria for a measure of scientific output. *Frontiers in Computational Neuroscience, 5*(48). doi: 10.3389/fncom.2011.00048.

Mahdi, S., d'Este, P., & Neely, A. D. (2008). *Citation counts: are they good predictors of RAE scores? A bibliometric analysis of RAE 2001*. London, UK: Advanced Institute of Management Research.

Mantel, N., & Haenszel, W. (1959). Statistical Aspects of the Analysis of Data from Retrospective Studies of Disease. *Journal of the National Cancer Institute, 22*(4), 719-748.

McKay, S. (2012). Social policy excellence - peer review or metrics? Analyzing the 2008 Research Assessment Exercise in social work and social policy and administration. *Social Policy & Administration, 46*(5), 526-543. doi: 10.1111/j.1467-9515.2011.00824.x.

National Information Standards Organization. (2016). *Outputs of the NISO Alternative Assessment Metrics Project*. Baltimore, MD, USA: National Information Standards Organization (NISO).

Neylon, C. (2014). Altmetrics: What are they good for? Retrieved October 6, 2014, from http://blogs.plos.org/opens/2014/10/03/altmetrics-what-are-they-good-for/#.VC8WETI0JAM.twitter

Plackett, R. L. (1974). *The analysis of categorical data*. London, UK: Chapman.

Radhakrishna, S. (1965). Combination of results from several 2 x 2 contingency tables. *Biometrics, 21*, 86-98.

Rons, N. (2012). Partition-based Field Normalization: An approach to highly specialized publication records. *Journal of Informetrics, 6*(1), 1-10. doi: 10.1016/j.joi.2011.09.008.

Rons, N. (2014). Investigation of Partition Cells as a Structural Basis Suitable for Assessments of Individual Scientists. In P. Wouters (Ed.), *Proceedings of the science and technology indicators conference 2014 Leiden "Context Counts: Pathways to Master Big and Little Data"* (pp. 463-472). Leider, the Netherlands: University of Leiden.

Sheskin, D. (2007). *Handbook of parametric and nonparametric statistical procedures* (4th ed.). Boca Raton, FL, USA: Chapman & Hall/CRC.

Smith, A., & Eysenck, M. (2002). *The correlation between RAE ratings and citation counts in psychology*. London: Department of Psychology, Royal Holloway, University of London, UK.

Thelwall, M. (2016). Three practical field normalised alternative indicator formulae for research evaluation. Retrieved from https://arxiv.org/abs/1612.01431

Thelwall, M. (2017a). Three practical field normalised alternative indicator formulae for research evaluation. *Journal of Informetrics, 11*(1), 128-151. doi: http://dx.doi.org/10.1016/j.joi.2016.12.002.

Thelwall, M. (2017b). *Web Indicators for Research Evaluation: A Practical Guide*. London, UK: Morgan & Claypool.

Thelwall, M., Kousha, K., Dinsmore, A., & Dolby, K. (2016). Alternative metric indicators for funding scheme evaluations. *Aslib Journal of Information Management, 68*(1), 2-18. doi: doi:10.1108/AJIM-09-2015-0146.

Waltman, L., & Costas, R. (2014). F1000 Recommendations as a Potential New Data Source for Research Evaluation: A Comparison With Citations. *Journal of the Association for Information Science and Technology, 65*(3), 433-445. doi: 10.1002/asi.23040.

Waltman, L., van Eck, N. J., van Leeuwen, T. N., Visser, M. S., & van Raan, A. F. J. (2011). Towards a new crown indicator: an empirical analysis. *Scientometrics, 87*(3), 467-481. doi: 10.1007/s11192-011-0354-5.

Williams, R., & Bornmann, L. (2016). Sampling issues in bibliometric analysis. *Journal of Informetrics, 10*(4), 1253-1257.

Work, S., Haustein, S., Bowman, T. D., & Larivière, V. (2015). *Social Media in Scholarly Communication. A Review of the Literature and Empirical Analysis of Twitter Use by SSHRC Doctoral Award Recipients*. Montreal, Canada: Canada Research Chair on the Transformations of Scholarly Communication, University of Montreal.

Wouters, P., Thelwall, M., Kousha, K., Waltman, L., de Rijcke, S., Rushforth, A., & Franssen, T. (2015). *The Metric Tide: Correlation analysis of REF2014 scores and metrics (Supplementary Report II to the Independent Review of the Role of Metrics in Research Assessment and Management)*. London, UK: Higher Education Funding Council for England (HEFCE).